\documentclass [12pt]{article}
\usepackage {amssymb}
\usepackage {amsmath}
\usepackage {graphicx}

\begin{document}

\footnotesize{\noindent  V. Krasnoholovets, Dark matter as seen from the physical point of view,  \textit{Astrophysics and Space Science}   {\bf 335}, No. 2, 619-627 (2011).}

\bigskip

\begin{center}
\Large {\bf Dark matter as seen from the physical point of view}
\end{center}

\vspace{3mm}

\begin{center}
{\bf Volodymyr Krasnoholovets}
\end{center}

\begin{center}
Indra Scientific SA, Square du Solbosch 26, Brussels B-1050, Belgium  \\
e-mail: {v\_kras@yahoo.com}
\end{center}

\bigskip

\begin{flushright}
7 May 2011
\end{flushright}

\abstract{
It is shown that the Newton's law of universal gravitation can be derived
from first submicroscopic principles inherent in the very nature of real
space that is constituted as a tessel-lattice of primary topological balls.
The submicroscopic concept determines the notion of mass in the
tessel-lattice and introduces excitations of space, which appear at the
motion of particles (mass particles are determined as local deformations of
the tessel-lattice). These excitations are associated with carriers of the
field of inertia. In the universe the gravitation is induced by standing
inerton waves of mass objects, which oscillate around the objects with the
speed of light. An overlapping of these standing inerton waves generates an
elastic interaction between masses bringing them to a formation of clusters
in which masses are characterised by both the Newtonian and elastic
interaction. It is this elastic interaction that cancels the necessity of
introduction of mystical dark matter. At the same time, inertons, 
carriers of inert properties of objects, can be treated as an analogous of  
hypothetic weakly interacting massive particles (WIMP) or axions, 
which some astronomers try to associate with dark matter particles.}

\bigskip

\textbf{Key words:} dark matter, clusters of galaxies, gravitation, space, tessel-lattice, inertons

\medskip

\textbf{PACS:} 04.20.Cv; 04.40.-b; 05.50.+q; 95.35.+d; 98.20.-d; 98.65.-r

\bigskip

\section{Introduction}

McGauph \cite{1} has recently reported results of testing the baryonic
Tully-Fisher relation (BTFR) for gas rich galaxies, i.e. for a class of
galaxies where stars do not dominate the baryonic mass budget. At the
testing both axes of the BTFR have been measured independently of the
available theories. McGauph has demonstrated that his data coincide exactly
with those predicted by Milgrom's hypothesis \cite{2,3}. Some
more theoretical preference to the Milgrom's modified Newtonian dynamics
(MOND) is given in a review article \cite{4}.

The starting point \cite{2,3} was a departure from Newtonian
dynamics. Namely, the Newton force ${\rm {\bf F}}=m{\kern 1pt}{\rm {\bf a}}$
was substituted for a significantly less force ${\rm {\bf F}}=m\cdot {\kern
1pt}({\kern 1pt}{\rm {\bf a}}-\Delta {\rm {\bf a}})$, which assumed to be
active at a scale of small centripetal accelerations of stars in some
galaxies. The modified regime is switched in at $a<<a_0 $ where $a_0
=(1.21\pm 0.24)\times 10^{-10}$ $\rm m{\kern 1pt}s^{-2}$ is the value of the threshould
acceleration estimated in paper \cite{5}. It is a matter of fact that
at such small $a$ the rotation curves of disc galaxies behave as flat:
stars' orbital velocities $V\sim \mbox{const}$, although in the classical
case of Newton dynamics the curves have to follow a Keplerian law, $V\sim
1/\sqrt r $.

Milgrom's approach allowed one to introduce a relation between a total mass
$M$ of the galaxy studied and its rotation velocity \cite{2,5}
\begin{equation}
\label{eq1}
V^4=GMa_0
\end{equation}

\noindent
and it is this relationship that has recently been verified \cite{1}.

Moreover, McGuaph \cite{1} further noted that MOND (i) predicted in advance
that galaxies of both high and low surface brightness would fall on the same
BTFR, though this contradicts to the expectation of purely Newtonian
gravity; (ii) prescribed the mass-to-light ratios that agreed the stellar
population synthesis models; (iii) provided the only successful a priori
prediction of the first-to-second peak amplitude ratio of the acoustic peaks
of the cosmic background radiation. McGuaph then reasonably concludes: ``It
is rare for a non-canonical theory to have so many predictive successes''.

Notwithstanding this, \cite{1} indicates also major shortcomings of MOND when
applying to describe rich clusters of galaxies \cite{6} and the
bullet cluster \cite{7}: the discrepancy of the clusters mass is
a factor of two, which means that MOND suffers of some kind of dark mass too
(though in the case of the Lambda-Cold Dark Matter model the discrepancy of
mass reaches 99\%).

Thus, to account for all the experimental results, MOND requires some dark
matter particles. These dark matter particles should behave themselves as
``standard cold dark'' matter on large scales and at the same time have to
interact with conventional matter resulting in MOND phenomenology for disc
galaxies.

In the present paper those missing dark matter particles are introduced
starting from first submicroscopic principles.

\section{Preliminaries}

\subsection{Mathematical background}

Some researchers mention the incompleteness of general relativity and the
difficulties associated with its physical interpretation. The others talk
about conceptual difficulties of quantum mechanics. And everyone understands
that these two theories are quite different and unification is possible only
with substantial changes in both concepts. Conciliation is feasible only on
a common physical base, which is beyond the formalism of both quantum
mechanics and general relativity. This would be rather a kind of the double
solution theory over which de Broglie \cite{8} was working since the beginning
of 1950-s.

So, what would be the starting point, a basis for a more universal physical
theory? The theory of everything tries to describe the whole physical world
starting from the first principles of quantum theory. However, what about a
theory of something? It is this theory that is able to clarify the major
primary notions of physics, namely: space, mass, charge, particle, field,
etc.

The theory of something \cite{9,10,11,12} starts as a
pure mathematical theory from set theory, topology and fractal geometry
generating a common study of such fundamental notions, as measure, distance,
space dimensions and the founding element (the empty set $\varnothing$). 
The axiom of the existence of the empty set, added with the axiom of availability, 
in turn provide existence to a lattice ${\cal L}(\varnothing)$ of empty set, 
which constitutes a discrete fractal space. The set of parts of $\varnothing$ 
contains parts equipotent to sets of integers, of rational and of real numbers, 
and owns the power of continuum.
These spaces provide collections of discrete manifolds whose interior is
endowed with the power of continuum. Any of intersections of subspaces
provide a (D$<$4)-space in which closed members get the status of both
observable objects and perceiving objects. This stands for observability,
which is a condition for a space to be in some sort observable, that is
physical-like.

Therefore, this mathematical lattice of empty set cells to be able to
account for a primary degenerate substrate \cite{9,12}. 
Space-time is represented by ordered sequences of topologically
closed Poincar\'{e} sections of this primary space. These mappings are
constrained to provide homeomorphic structures serving as frames of
reference in order to account for the successive positions of any objects
present in the system. Discrete properties of the lattice, called a
\textit{tessel-lattice}, allow the prediction of scales at which microscopic
 to cosmic structures should occur.

Deformations of primary cells by exchange of empty set cells allow a cell to
be mapped into an image cell in the next section as far as mapped cells
remain homeomorphic. However, if a deformation involves a fractal
transformation to objects, there occurs a change in the dimension of the
cell and the homeomorphism is not conserved. Then the fractal kernel of such
deformed cell of the tessel-lattice can stand for a ``particle'' and the
reduction of its volume (together with an increase of its area) is
compensated by morphic changes of a finite number of surrounding cells.

It is obvious that in the tessel-lattice a moving particle-like deformation
has to interact with the surrounding cells involving a fractal decomposition
process: the particulate cell exchanges its original deformation with the
surrounding cells, which will result in the appearance of a kind of a cloud
of deformed cells enclosing the particulate cell.

\subsection{The emergence of quantum physics in the tessel-lattice}

A volumetric fractal deformation of a cell of the tessel-lattice can be
associated with the physical notion of mass, 
$m=C{\cal V}_{\rm deg. {\kern 1pt}cell}/{\cal V}_{\rm deform. {\kern 1pt}cell}$, where
$C$ is a dimension constant and ${\cal V}_{\rm deg. {\kern 1pt}cell}$ stands
for the volume of an original degenerate cell and ${\cal V}_{\rm deform. {\kern 1pt}cell}$ for
the volume of the deformed cell. In physics a resistance to the motion is
called inertia. That is why excitations of the tessel-lattice produced by a
moving particulate cell were called \textit{inertons}.

If we consider kinetics of the motion of such complex object -- a particle
surrounded with a cloud of inertons, -- we derive \cite{13,14} 
relationships for a particle suggested by Louis de Broglie in
1924 
\begin{equation}
\label{eq2}
E=h {\kern 1pt} \nu,
\quad
\lambda = h / (mv).
\end{equation}

Since relationships (\ref{eq2}) have been obtained for a particle moving in the
tessel-lattice, the sense of parameters in (\ref{eq2}) becomes very clear: $E$ is 
the total energy of the particle, $mv$ is its momentum, $\lambda$ is 
the spatial period in which the particle emits inertons and then absorb them back, 
$\nu$ is the frequency of collisions of the particle with its cloud of inertons. Because
of the periodicity, the moving particle periodically emits and absorbs its
inertons, it is characterised by an increase in its action $\Delta S$; 
since this is a free motion through the degenerate tessel-lattice, this increment 
$\Delta S$ is associated with the Planck constant $h$.

De Broglie's relationships (\ref{eq2}) allow one to obtain the Schr\"{o}dinger wave
equation \cite{15}. But the relationships (\ref{eq2}) obtained in our case
signify that the Schr\"{o}dinger wave $\psi$-function gains a
real physical sense: $\psi$ represents the particle's field of
inertia whose carriers are inertons and they carry mass and fractal
properties of the particle \cite{16}. The cloud of inertons
occupies the section $\lambda$ along the particle's path and
spreads to a distance of $\Lambda = \lambda c/v$ in transversal directions,
where $c$ is the velocity of light for inertons in the
particle's cloud (though a free inerton possesses a large velocity \cite{17}).

An inerton field complete rejects the action at-a-distance from the realm of
quantum physics and introduces determinism at every stage of evolution of the
system studied \cite{16}. This new submicroscopic concept could
properly resolve quite a number of difficulties available in quantum
mechanics. Moreover, the concept was successfully applied for the
description of many experiments. Most interesting applications, which are in
line with the subject of the present paper, are associated with a variation in
mass of entities in condensed media (owing to the overlapping of inerton
clouds of entities) \cite{18} and the dense inerton field
gathering of about $10^{10}$ electrons in a log-living cluster
(due to the absorption of inertons knocked out by a laser beam from a
crystal) \cite{19}.

The tessel-lattice is the source for an electric and magnetic charge, photon
and electromagnetic field and these make it possible to derive the Maxwell
equations \cite{16}.

\subsection{Gravitation in the tessel-lattice}

At last, the concept allows the derivation of the Newton's law of universal
gravitation \cite{20,16}, which is behind the formalism of
general relativity. In the case of a canonical particle, the gravitation
emerges when the tessel-lattice responses to inertons emitted by the moving
particle, namely, the tessel-lattice elastically pushes them back to the particle
and the latter re-absorbs them. This means the motion of a particle occurs
with the periodical decay of its mass: the mass is gradually transformed
into a tension of the tessel-lattice. The appropriate whole Lagrangian has
the form \cite{20}
\begin{equation}
\label{eq3}
\begin{array}{l}
L=-m_0c^2 \Big\{ {\rm submicroscopic} {\kern 3pt} {\rm mechanics} \quad\quad\quad   \\
\qquad  + \Big(\dfrac{T^2}{2m_0^2} {\kern 1pt} {\dot m}^2 + \dfrac{T^2}{2\Lambda^2} 
  {\kern 1pt} {\dot {\boldsymbol \xi}} ^{{\kern 1pt}2}
- \dfrac{T}{m_0}  {\kern 1pt} {\dot m} \nabla {\boldsymbol \xi}  {\kern 1pt} \Big)\Big\}^{1/2}. 
\end{array}
\end{equation}

Here, $m_0$ is the rest mass of the particle, $\Lambda$ is the amplitude of 
the particle's inerton cloud, $T$ is the period of collisions of the particle and its inerton
cloud; $m=m(\boldsymbol r, {\kern 2 pt}t)$ is the current mass of the {\{}particle + inerton
cloud{\}}-system; $\boldsymbol \xi= \boldsymbol \xi(\boldsymbol r, {\kern 2pt}t)$ 
is the current value of the rugosity of the tessel-lattice in the range covered by the system. 
Geometrically the rugosity $\boldsymbol \xi$ depicts the state in which the tessel-lattice
cells covered by the inerton cloud are shifted a little bit from their
equilibrium positions, which induces a local tension in the tessel-lattice;
the tension of the particulate cell may be associated with an increase in
size.

The system studied features the radial symmetry; then variables $m$ and $\boldsymbol \xi$ 
are functions of only the distance $r$ from the particle and the proper time $t$ of the
{\{}particle + inerton cloud{\}}-system. In this case we preserve only
radial components in the both variables, which enables us to use the
spherical coordinates. The Euler-Lagrange equations of motions for $m$ and the radial component 
of $\boldsymbol \xi$ result in \cite{20}
\begin{equation}
\label{eq4}
\frac{\partial^2 m}{\partial {\kern 1pt} t^2} -\Big( \frac{c^2}{r}\Big)
\frac{\partial^2 (r {\kern 1pt} m)}{\partial {\kern 1pt} r^2}=0, \quad
\end{equation}
\begin{equation}
\label{eq5}
\frac{\partial^2 \xi}{\partial {\kern 1pt} t^2} -\Big( \frac{\Lambda^2}{m_0T}\Big)
\frac{\partial}{\partial {\kern 1pt} r}
\Big(\frac{\partial {\kern 1pt} m}{\partial {\kern 1pt} t}\Big)=0
\end{equation}
where the Laplace operator is presented in the spherical coordinates as
$\Delta m =\frac{1}{r}\frac{\partial^2}{\partial r^2}(rm)$.

The radial symmetry allows the solutions to equations (\ref{eq4}) and (\ref{eq5}) in the
form of standing spherical waves, which exhibit the dependence $1/r$,
\begin{equation}
\label{eq6}
m(r,{\kern 2 pt}t )=C_1\frac{m_0}{r}\cos \Big(\frac{\pi r}{2\Lambda}\Big) \Big | \cos \Big( \frac{\pi t}{2T}
\Big) \Big |, \quad\quad
\end{equation}
\begin{equation}
\label{eq7}
\xi(r,{\kern 2 pt}t )=C_2\frac{\xi_0}{r} \sin \Big( \frac{\pi r}{2\Lambda}\Big) 
\cdot  \big( -1 \big)^ { \Big[ \dfrac{t} {T}\Big] } \cdot \Big | \sin\Big( \frac{\pi t}{2T}\Big) \Big |.
\end{equation}
The dimensionality of integration constants $C_{1,{\kern 1pt}2}$ corresponds
to length and can be put here $C_1=l_{\rm Planck} \approx 10^{-35}$ m and $C_2=\Lambda$.

An object, which consists of many particles (a solid, a planet, or a star),
experiences local vibrations of its entities (atoms, ions, particles), as is
the case with entities in the crystal lattice. Entities vibrate in the
neighborhood of their equilibrium positions and/or move to new positions.
These movements produce inerton clouds around the appropriate particles.
Inerton clouds overlap forming a total inerton cloud of the object
\cite{20,16}. The spectrum of inertons is similar to the
spectrum of phonons in a solid (a body of phonons is filled with inerton
carriers) \cite{18}. For instance, if we have a solid sphere
with a radius $R_{\rm sph}$, which consists of $N_{\rm sph}$
atoms, the spectrum of acoustic waves will include $N_{\rm sph}/2$ wave
harmonics with wavelengths $\lambda_n=2bn$ where $b$ is
the lattice constant and $n=1,{\kern 2pt}2,{\kern 2pt}3,{\kern 2pt}...,{\kern 2pt} N_{\rm sph}/2$.

At the same time, overlapping inerton clouds, which accompany vibrating
atoms, produce their own spectrum of inerton wavelengths
\begin{equation}
\label{eq8}
\Lambda_n =\lambda_n {\kern 1pt} c/v_{\rm sound}.
\end{equation}
This means that the Lagrangian (\ref{eq3}) and the solutions (\ref{eq6}) and (\ref{eq7}) obtained
for a moving particle are proper also for a macroscopic object at rest.
Indeed, an object with a mass $m$ is characterised by the
inner motion of its entities, whose vibrations induce oscillations of an
inerton field both inside and outside of the object.

For instance, for a small solid sphere with a size 1 cm$^{3}$ and the
quantity of atoms $N_{\rm sph} \sim 10^{22}$ we get from relation (\ref{eq8}) that the
longest standing inerton wave can spread up to a distance $\Lambda \sim 10^{17}$ m 
$\approx 10.5$ light years. Behind this radius any information about the solid sphere is absent.

Longest standing inerton waves of a macroscopic body with a radius $R$ and a mass $m$ 
induce the deformation potential (\ref{eq6}) in the surrounding space, 
i.e. these waves contract cells of the tessel-lattice as the rule (\ref{eq6}) prescribes. 
In the region of space $R \le r << \Lambda$ the time-averaged distribution of the mass of the standing
inerton wave becomes
\begin{equation}
\label{eq9}
m(r, {\kern 2pt} t) = l_{\rm {\kern 1pt} Planck}{\kern 2pt} m_0/r.
\end{equation}
Multiplying both hand sides of expression (\ref{eq9}) by a factor of $-G/l_{\rm Planck}$ 
we obtain a conventional Newtonian potential
\begin{equation}
\label{eq10}
U=-Gm_0/r
\end{equation}
that describes the gravitational attraction of a test mass to the object
with the mass $m_0$.

This result shows that the gravitational mass of the object is complete
accumulated in its standing inerton wave.

This theory of the Newtonian potential formation has further been developed
in works \cite{21,22}. In particular, it has been argued the
necessity of the tangential inerton interaction between masses, which gives
rise to a correction to the Newton's law of gravitation \cite{21}
\begin{equation}
\label{eq11}
U=-G {\kern 1pt} \frac{m_0 {\kern 1pt} m_1}{r} \Big(1+\frac{\dot r^2_{\rm tan}}{c^2}\Big)
\end{equation}
where $\dot r_{\rm tan}$ is the tangential velocity of a test body with the
mass $m_1$ i.e. the body's orbital velocity. Corrected
Newton's law of gravitation (\ref{eq11}) was further used to study the anomalous
precession of the Mercury's perihelion, the bending of light and the red
shift of spectral lines \cite{21}. The submicroscopic concept
made it possible to derive exactly the same equations for the descriptions
of those three phenomena, which were produced by general relativity.
Moreover, the submicroscopic concept has enabled one to clarify a nature of
changes in space associated with the so-called gravitational time delay
effect (the Shapiro time delay effect) \cite{22}. It is
important to note that the velocity of inertons in standing inerton waves
may exceed the speed of light, because their velocity includes also the
velocity of the object that irradiates these standing waves (i.e., 
$v_{\rm stand. {\kern 2pt} inertons}=\sqrt{c^{ {\kern 0.5pt} 2} 
+ v^{{\kern 0.5pt}2}_{\rm {\kern 0.5pt} object}}$, 
see the expression (\ref{eq11}) and Ref. \cite{21}).

\section{Interacting stars}

\subsection{Statistical mechanical approach}

Krasnoholovets and Lev \cite{23} developed a method of statistical physics for
the description of systems of interacting particles taking into account a
spatial nonhomogeneous distribution of particles, i.e. cluster formation. In
particular, we considered gravitating masses with the Hubble expansion.

Identical gravitating masses, i.e. stars, were characterised by an
effective attraction potential energy $u_{12}^{\rm attr.}=Gm^2/(r_1-r_2)$ and the effective
repulsion energy $u_{12}^{\rm repul.}=\frac{1}{2}mH_0^2\cdot (r_1-r_2)^2$. 
We used the fact that an additional kinetic energy $E_{12}=\frac{1}{2}m\cdot(v_1-v_2)^2$
 of masses is associated with the Hubble expansion. The relative velocity~$v_1-v_2$ of 
 masses separated in space correlates with the relative distance between the masses, because
$v_1-v_2=H_0 \cdot (r_1-r_2)$,~where $H_0$ is the Hubble constant.

We obtained the following solution for a number of stars gathered in a
cluster \cite{23}
\begin{equation}
\label{eq12}
\aleph \approx \frac{120 {\kern 1pt}  \pi}{33} {\kern 1pt} \frac{G\rho}{H_0^2}
\end{equation}
where $\rho$ is the density of stars. As seen from expression
(\ref{eq12}), just value of $\rho$ is critical for the formation of a
cluster, i.e. when $\aleph >> 1$. Putting $H_0 =1.6 \times 10^{18}$
s$^{-1}$, we may state that the inequality $\aleph >>1$ holds when
the density of stars $\rho>> 10^{-25}$ kg m$^{-3}$ 
$\approx {\rm M}_\odot /(11{\kern 2pt} {\rm kpc})^3$.

Now let us come back to the Newtonian gravitational potential (\ref{eq10}), which is
formed by standing inerton waves of an object with a mass $m$. 
In the case of the sun, the number of its particles is above
$10^{50}$. Then at the worse conditions for relation (\ref{eq8}), $\lambda_1 \sim 10^{-15}$ m 
and $c/v_{\rm sound} \sim 1$, we get $\lambda_{N/2} \sim 10^{35}$ m and may
crude estimate a boundary to which the sun's inertons can spread: $\Lambda_{N/2} \sim 10^{35}$ m, 
which exceeds the observed radius of the universe. This means that we may neglect 
the rugosity/tension (\ref{eq7}) of the tessel-lattice at an examination of a system of stars, 
which are disposed in the same galaxy.

However, we cannot disregard an overlapping of local deformations of the
tessel-lattice, induced by standing inerton waves of a system of stars. In
fact, a ``breathing'' of a star, i.e. radial oscillations of its inerton
clouds, which occur with the speed of light $c$, results in
mutual overlapping of inerton clouds of stars. Therefore, these standing
inerton waves induce the Newton's potential of gravitation (10) and, in addition, 
owing to the mutual scattering of counterpropagating waves of nearest stars
they introduce the elasticity in interstellar space. 
This means that the quasi-stationary gravitational law (\ref{eq11}) should be supplemented 
by an additional elastic energy created by the mutual overlapping of inerton
clouds of all stars of the system studied. Note in such a way a unification
of molecules takes place in gases, liquids and solids \cite{18}.

Thus, a correct expression for the energy of interacting stars should
include four terms: (i) the gravitational potential interaction (\ref{eq11}) between
two masses $m_i$ and $m_j$; (ii) the gravitational interaction of a mass $m_i$ 
with the total mass $M$ of the system of stars; (iii) an elastic interaction
between masses $m_i$ and $m_j$; (iv) the interaction between masses 
$m_i$ and $m_j$ associated with the Hubble expansion (see above).

Basing on the results \cite{23}, we may assume that all
stars in the system studied are distributed by nodes of a lattice (some
nodes are filled and some not). Then if a system of interacting particles
possesses attraction and repulsion pair potentials, statistical mechanics
prescribes \cite{23} that in such a system all particles
become distributed by $K$ identical clusters. The action for
each cluster looks as follows
\begin{equation}
\label{eq13}
S \approx (\alpha - \beta)\cdot \aleph^2
\end{equation}
where $\aleph$ is the number of particles in the cluster and $\alpha$ and $\beta$ 
are functions associated with the particle interactions:
\begin{equation}
\label{eq14}
\begin{array}{l}
\alpha=\dfrac{1}{\wp {\kern 1pt} k_{\rm B} \Theta}\int d {\kern 0.5pt} {\boldsymbol r} {\kern 2pt} 
u^{\rm elast. {\kern 2pt} repul.} ({\boldsymbol r}), \\
\beta=\dfrac{1}{\wp {\kern 1pt} k_{\rm B} \Theta}\int d {\kern 0.5pt} {\boldsymbol r} {\kern 2pt} u^{\rm attract.} 
({\boldsymbol r})\\
\end{array}
\end{equation}
where $\wp$ is the effective volume of a particle and the integration is running over 
the volume of the whole cluster; $k_{\rm B} \Theta$ is the thermal energy of the environmental thermostat
($k_{\rm B}$ is the Boltzmann constant and $\Theta$ the
absolute temperature). Absolute values of the pair potential interactions
are
\begin{equation}
\label{eq15}
u^{\rm elast.{\kern 1pt} repuls.} (gx) = \frac{1}{2} {\kern 1pt} m {\kern 1pt} \omega^2 \cdot (g {\kern 1pt} x)^2 
+\frac{1}{2} {\kern 1pt} m {\kern 1pt} H_0^2 \cdot (g {\kern 1pt} x)^2,
\end{equation}
\begin{equation}
\label{eq16}
u^{\rm attract.}(gx)=\frac{GMm}{R} +\frac{Gm^2}{gx} \qquad\qquad\qquad\qquad
\end{equation}
where $m$ is the mass of a star; $g$ is the lattice constant 
(a distance between neighbour stars in the model lattice);
$x$ is the dimensionless distance defined through a relation
$r=gx$; $\omega$ is the radial frequency of
oscillation of the mass $m$ near its equilibrium position;
$M$ is the mass of the whole system of stars, which occupies
space up to the effective radius $R$, and hence in
expression (\ref{eq16}) the first term can be considered as a middle-field potential
energy.

\subsection{3-D clusters}

For a spherical cluster the integrals in expression (\ref{eq14}) can be rewritten
via the number $\aleph$ of particles in the cluster \cite{23}

\begin{equation}
\label{eq17}
\frac{1}{\wp}\int d {\kern 0.5pt} \boldsymbol r = \frac{1}{(4\pi/ 3)g^3} 
{\kern 1pt} 4\pi \int_g^R r^2 d {\kern 0.5pt} r = \frac{R^3-g^3}{g^3} = \aleph .
\end{equation}

Substituting expressions (\ref{eq15}) and (\ref{eq16}) into the integrals (\ref{eq14}) we get
\begin{equation}
\label{eq18}
\begin{array}{l}
\alpha =\dfrac{3}{k_{\rm B}\Theta} \int_1^{\aleph^{1/3}} d {\kern 0.5pt} x{\kern 1pt}x^2 u^{\rm elast.{\kern 2pt} repuls.}(gx)    \\
\quad  = \dfrac{3}{10 {\kern 1pt} {k_{\rm B}\Theta}} {\kern 1pt} mg^2\cdot (\omega^2+H_0^2) {\kern 1pt}\aleph^{5/3} ,
\end{array}
\end{equation}
\begin{equation}
\label{eq19}
\begin{array}{l}
\beta = \dfrac{3}{k_{\rm B}\Theta} \int_1^{\aleph^{1/3}} d {\kern 0.5pt} x{\kern 1pt}x^2 u^{\rm attract.} (gx)   \\
\quad  = \dfrac{3}{k_{\rm B}\Theta}\Big( \dfrac{GMm}{3R}\aleph + \dfrac{Gm^2}{2g}\aleph^{2/3} \Big).
\end{array}
\end{equation}
Then the action (\ref{eq13}) becomes
\begin{equation}
\label{eq20}
\begin{array}{l}
S= \dfrac{1}{k_{\rm B}\Theta}\Big\{ \dfrac{3}{10} {\kern 1pt} m \cdot (\omega^2+H_0^2){\kern 1pt}  
g^2 {\kern 1pt} \aleph^{11/3}      \\
\qquad \qquad \quad  - \dfrac{GMm}{R}\aleph^3 - \dfrac{3{\kern 1pt}G  m^2}{2g}\aleph^{8/3} \Big\}.
\end{array}
\end{equation}

Taking into account the fact that galaxies and clusters consist at least of
a few thousand stars and knowing typical values for $M$,
$m$, $R$ and $g$, we may conclude that in the first approximation 
the last term in expression (\ref{eq20}) can be neglected. 
Besides, it is reasonable to assume that the Hubble
energy, which affects a star, is smaller than the elastic energy that
retaining the star in a cluster. Therefore, the contribution on the side of
$H_0$ may be considered as negligible. The simplified action
(\ref{eq20}) can be investigated for the extremum: $\partial S / \partial \aleph =0$. 
The solution to this equation is
\begin{equation}
\label{eq21}
\aleph = \Big( \frac{30}{11}\frac{GM}{Rg^2\omega^2} \Big)^{3/2} .
\end{equation}

\subsection{2-D clusters}

In the case of quasi-flat clusters the expression (\ref{eq17}) changes to
\begin{equation}
\label{eq22}
\frac{1}{A} \int d {\kern 0.5pt} \boldsymbol r = \frac{1}{\pi g^2} {\kern 2pt} 2\pi
\int_g^R r{\kern 1pt} d {\kern 0.5pt} r =
\frac{R^2 - g^2}{g^2} = \aleph
\end{equation}
where $A=\pi g^2$ is the area occupied by one particle in a cluster.
Then retaining highest order terms in equation (\ref{eq14}) written for the flat
case, we obtain for the functions $\alpha$ and $\beta$:
\begin{equation}
\label{eq23}
\alpha = \frac{2}{k_{\rm B}\Theta} \int_1^{\aleph^{1/2}}d {\kern 0.5pt} x {\kern 1pt} x {\kern 1pt} 
u^{\rm elast. {\kern 2pt}repuls.} (gx) = \frac{1}{2}\frac{m {\kern 1pt} \omega^2 g^2}{k_{\rm B}\Theta} 
\aleph^2 , 
\end{equation}
\begin{equation}
\label{eq24}
\beta = \frac{2}{k_{\rm B}\Theta} \int_1^{\aleph^{1/2}} d {\kern 0.5pt} x{\kern 1pt} x{\kern 1pt} 
u^{\rm attract.} (gx) =\frac{GMm}{R {\kern 1pt} k_{\rm B}\Theta}.\qquad\quad\quad
\end{equation}
Having these functions, we construct the action (\ref{eq13}) as follows
\begin{equation}
\label{eq25}
S = \frac{1}{k_{\rm B}\Theta} \Big \{ \frac12 m {\kern 1pt} \omega^2 g^2 {\kern 1pt} \aleph^4 
-\frac{GMm}{R} {\kern 1pt} \aleph^3 \Big\}.
\end{equation}
The extremum is achieved at the solution of the equation $\partial S / \partial \aleph =0$, 
which results in the solution
\begin{equation}
\label{eq26}
\aleph = \frac32 \frac{GM}{R{\kern 1pt}\omega^2g^2} .
\end{equation}

\subsection{1-D clusters}

In the case of quasi-linear clusters formulas are maximally simplified.
Indeed, $\aleph=R/g$ and the functions $\alpha$ and $\beta$ (\ref{eq14}) become
\begin{equation}
\label{eq27}
\alpha =\frac{1}{k_{\rm B}\Theta} \int_1^\aleph d {\kern 0.5pt} x {\kern 1pt} u^{\rm elast. {\kern 2pt}{repuls.}}
(gx) = \frac13 \frac{m {\kern 1pt} \omega^2 g^2}{k_{\rm B}\Theta} {\kern 1pt} \aleph^3,
\end{equation}
\begin{equation}
\label{eq28}
\beta = \frac{1}{k_{\rm B}\Theta} \int_1^\aleph d {\kern 0.5pt} x {\kern 1pt} u^{\rm attract.} (gx) 
=\frac{GMm}{R{\kern 1pt} k_{\rm B}\Theta} {\kern 1pt} \aleph.  \qquad\quad
\end{equation}
Having functions (\ref{eq27}) and (\ref{eq28}), we obtain the action (\ref{eq13}) as follows
\begin{equation}
\label{eq29}
S = \frac{1}{k_{\rm B}\Theta} \Big\{ \frac13 m {\kern 1pt} \omega^2 g^2{\kern 1pt} \aleph^5 
-\frac{GMm}{R} {\kern 1pt}  \aleph^3 \Big\} 
\end{equation}
The solution of the equation $\partial S / \partial \aleph =0$ results in
\begin{equation}
\label{eq30}
\aleph = \Big( \frac95 \frac{GM}{R {\kern 1pt} \omega^2 g^2}\Big)^{1/2}.
\end{equation}

Let us analyse the obtained solutions (\ref{eq21}), (\ref{eq26}) and (\ref{eq30}).

\section{Discussion}

The solutions, which exhibit the distribution of particles by clusters with
$\aleph$ particles per cluster, can be applied for the
description of disc galaxies (expression (\ref{eq26})) and star clusters (expression
(\ref{eq21})). In fact, the phenomenon of cluster formation is well known in
condensed matter physics, which occurs with the presence of an outside field. For
example, in the presence of a thermal gradient the so-called
Rayleigh-B\'{e}nard cells (identical cylindrical or hexagonal structures)
appear in a layer of a primary uniform viscous fluid, electrons assemble in
clusters (about $10^{8}$ electrons per cluster) on the surface of
liquid helium and, at last, we \cite{19}
could generate a long-living clusters of electrons in which about $10^{10}$ 
electrons were gathering in one droplet where they were hold by
an inerton field.

\begin{figure}
\begin{center}
\includegraphics[scale=0.7]{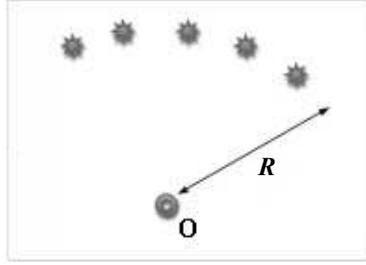}
\caption{\small Arc formed by stars in the neighbour of the centroid of a
galaxy, which is a typical quasi-1-D cluster.} \label{Figure 1}
\end{center}
\end{figure}

Arcs and arclets of stars have been observed and intensively investigated in
rich clusters of galaxies \cite{24}. 1-D cluster solution (\ref{eq30})
exactly satisfies an arc of stars (Fig. 1). Indeed, $\aleph >> 1$ is
reached in a wide range of parameters. For example, putting $M \sim 10^{13}{\kern 1pt}{\rm M}_\odot$, 
$R \sim 100$ kpc, $g \sim 1$ pc and $\omega \sim 10^{-14}$ s$^{-1}$, 
we get from (\ref{eq30}): $\aleph \sim 10^5$ stars in an
arc. Less values of $M$ will give less value of stars
involved in an arc. It is interesting that in principle a ring distribution
of stars is also quite possible.

For a disc galaxy we may apply the 2-D cluster solution (\ref{eq26}), which means
that all the stars of the disc galaxy are distributed by plane clusters with
the appropriate number $\aleph$ of stars. Let us estimate the value of $\aleph$. 
In particular, for a disc gas galaxy we may choose some
typical values of the mass, the radius and the distance between stars in the
galaxy: $M=10^8 {\kern 1pt} {\rm M}_\odot$ \cite{1}, $R=10$ kpc, $g=1$ pc. 
So, we have only one fit parameter, the frequency $\omega$ of oscillations 
of a star near its equilibrium position in the cluster. 
This parameter can be estimated from a relation that restrains a
star in the cluster. Figure 1 depicts: star 1 experiences the centripetal
acceleration \textbf{\textit{a}} to the centre O of the galaxy; at the same
time neighbour stars strongly keep it by means of the elastic energy 
$\frac12 m {\kern 1pt} \omega^2 {\boldsymbol r}^{{\kern 1pt }2}$ (Fig. 2). 
This means that the equality of two accelerations is hold:
\begin{equation}
\label{eq31}
GM/R^2 - \omega^2 g = 0.
\end{equation}
From equation (\ref{eq31}) we get

\begin{equation}
\label{eq32}
\begin{array}{l}
\omega = \Big( \dfrac{GM}{R^2g} \Big) \approx \Big\{ \dfrac{(6.67\times 10^{-11})
\times (2\times 10^{40})} {(3.09 \times 10^{20})^2 \times (3.13 \times 10^{16})} \Big\}^{1/2} \\
\qquad \qquad \quad  \simeq 2.1 \times 10^{-14} \ [\rm s^{-1}]. 
\end{array}
\end{equation}

\begin{figure}
\begin{center}
\includegraphics[scale=0.65]{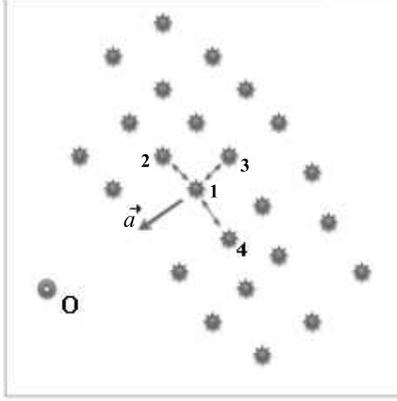}
\caption{\small Stars in a disc galaxy. The acceleration \textbf{\textit{a}},
which star 1 experiences to the centre O of the galaxy, competes
with accelerations on the side of surrounding stars 2, 3, 4, etc., which
restrain the motion of star 1 to the point O.} \label{Figure 2}
\end{center}
\end{figure}

Having known all the parameters, we may estimate the number of stars in a
cluster (\ref{eq26}): $\aleph \approx 1.5 \times 10^4$.

In this plane cluster each star is involved in three kinds of interactions
whose energies are
\begin{equation} 
\label{eq33}
\begin{array}{l}
GMm/R \approx 8.63 \times 10^{41} {\kern 2pt} {\rm J}, \\ 
Gm^2/g \approx 8.52 \times 10^{37} {\kern 2pt} {\rm J}, \\
\frac 12 m {\kern 1pt} \omega^2 g^2 \approx 4.32 \times 10^{36} {\kern 2pt} {\rm J}. \\
\end{array}
\end{equation}

Let us now evaluate the acceleration that each star experiences in the plane
cluster:

\begin{equation} 
\label{eq34}
a = GM/R^2 \approx 1.4 \times 10^{-11} \ {\rm m {\kern 2pt} s^{-2}}.
\end{equation}

This value of the acceleration satisfies the conditions prescribed by
\cite{2,3}: $a<< a_0 \simeq 1.21 \times 10^{-10}$ m s$^{-2}$. 
Thus we do not need to assume an incomprehensible modernization of Newton's law, 
i.e. the substitution of the force ${\bf F} =m{\bf a}$ by a significantly less force
${\bf F} = m \cdot ({\bf a} - \Delta {\bf a})$ at $a<< a_0$. Stars are distributed by
clusters and each star is strongly bounded with the other $\aleph -1$ cluster's stars. 
This bounding compensates the centripetal acceleration $a$, 
which directs stars to the centre of the gravitating potential of the total mass 
$M$ of the galaxy. Expression (\ref{eq31}) demonstrates this balance of two competing forces. That is
why a Keplerian law, $V \sim \sqrt {r}$, is substituted for the constant
orbital velocity (\ref{eq1}).

Let us discuss now the 3-D cluster solution (\ref{eq21}). The observed cluster (the
interacting cluster 1E 0657-558 \cite{7}) has the total mass
$M \sim 10^{14} {\kern 1pt} {\rm M}_\odot$, the radius $R \approx 250$ kpc and the central mass
density $\rho_0 = 3.85 \times 6 {\kern 1pt} {\rm M}_\odot$ kpc$^{-3}$. Putting for the mass of a star
$m = {\rm M}_\odot$, we obtain the mean distance between stars: 
$g=({\rm M}_\odot/\rho_0)^{1/3} = 4.25 \times 10^{16}$ m. 
Then the stability of the cluster in respect to its
gravitational collapse is determined by the relationship (\ref{eq31}): the
gravitational attraction of stars to the centroid is retained by the
elasticity of inerton waves in the cluster. The frequency of oscillating
stars at their equilibrium positions owing to the overlapping of their
inerton waves is

\begin{equation}
\label{eq35}
\begin{array}{l}
\omega = \Big( \dfrac{GM}{R^2g} \Big) \approx \Big\{ \dfrac{(6.67\times 10^{-11})
\times (10^{14}\times 2\times 10^{32})} 
{(7.71 \times 10^{21})^2 \times (4.25 \times 10^{16})} \Big\}^{1/2}  \\
\qquad\qquad\quad \   \   \simeq 3.82 \times 10^{-13} \ [\rm s^{-1}]. 
\end{array}
\end{equation}

The number of stars in such cluster, as it follows from expression (\ref{eq21}),
$\aleph = 2.39 \times 10^9$.

In this 3-D spherical cluster each star participates in three kinds of
interactions whose energies are

\begin{equation} 
\label{eq36}
\begin{array}{l}
GMm/R \approx 3.46 \times 10^{46} {\kern 2pt} {\rm J}, \\ 
Gm^2/g \approx 6.27 \times 10^{37} {\kern 2pt} {\rm J}, \\
\frac 12 m\omega^2 g^2 \approx 2.64 \times 10^{39} {\kern 2pt} {\rm J}. \\
\end{array}
\end{equation}

The acceleration to the centroid, which each star experiences in the
spherical cluster, is

\begin{equation} 
\label{eq37}
a = GM/R^2 \approx 2.24 \times 10^{-8} \ {\rm m {\kern 1pt} s^{-2}}.
\end{equation}

The acceleration (37) is opposite to the inequality $a<<a_0$
needed for the use of MOND \cite{2}. Besides, the acceleration (37)
is not compensated by the acceleration caused by the elastic interaction in
the cluster of $\aleph$ stars: $\omega^2 g \approx 6.2 \times 10^{-9}$ m s$^{-2}$.
These are the reasons why calculations \cite{7} of the shear
profile $\gamma (\theta)$ caused by a point mass, which included a
correction based on the MOND approach, showed a discrepancy between the
available mass and the too intensive X-rays. Clowe et al. \cite{7} note that
the dark matter in the cluster exceeds at least twice the baryonic mass
component in a MOND regime. The same emphasizes McGauph \cite{1}.

However, the origin of so-called dark matter is nothing but the same stars,
which are involved in the mutual interaction through their inerton waves.
This means that data obtained from the observation of stars must be
considered taking into account an inerton component bounding stars. An
important role may play parameters in expressions (33) and (36).

Basic concepts of gravitational lensing \cite{25,26,27} should also be modified -- perhaps a
point mass approach with a correction based on MOND or the other model
will require a substitution by an approach resting on the involvement of
elastically interacting masses. In particular, it seems the deflection angle
$\varphi =4Gm/(c^2 r)$ of a point mass $m$, which includes the
absolute value of the gravitational potential $Gm/r$, can be
modified as follows
\begin{equation}
\label{eq38}
\tilde \varphi = \frac{4}{c^2} \Big( \frac{Gm}{r} + \frac{1}{2} {\kern 1pt} \omega^2 {\kern 1pt} r^2 \Big);
\end{equation}
this is evident from the pair interactions of stars (\ref{eq15}) and (\ref{eq16}). 
In a cluster the second term in expression (38) tends to align the space deformed by the first term. 
This has to be typically for 2-D clusters (i.e. clusters in disc galaxies), which is apparent from 
expressions (33). In the case of 3-D clusters (rich clusters in galaxies) the second term may even 
prevail the first one, see expressions (36), namely, the second term prolongs the deflection 
angle $\varphi$  for larger distances at which the first Newton's term becomes already negligible. 
    
The deflection angle (\ref{eq38}) can be presented through a ratio of accelerations
\begin{equation}
\label{eq39}
\tilde \varphi = \frac{4Gm}{c^2 {\kern 1pt} r} \Big( 1 + \frac{a \quad}{2a_{\rm grav.} }\Big)
\end{equation}
where $a=\omega^2 {\kern 0.5pt} r$ and $a_{\rm grav.} = G {\kern 1pt} m/r^2$, 
which brings the approach closer to the MOND hypothesis. In a general way, 
one has to take into account a sum of distributed point masses and the possible 
presence of an outside potential $-GM/R$.

Massey et al. [28] showed that dark matter does interact via gravity, 
which is most effectively probed through gravitational lensing. 
The correction to the deflection angle introduced in expression (38) discloses the reason for 
such behaviour of dark matter. 

Maps of the large-scale distribution of dark matter, a network of filaments and their 
intersection revealed by Massey et al. [28] allow a reasonable interpretation in 
the framework of the present theory: standing inerton waves of large gravitating 
masses indeed must interfere quite similar to waves on the water surface. An evolution   
of such interference pattern tends to a peculiar gravitational background, or scaffold by Massey et al. [28], 
into which gas can accumulate, and stars can be built.

\section{Concluding remarks}

In the present work we have shown that the submicroscopic concept exhibits
the gravitation as a dynamic phenomenon -- no motion, no gravity, -- and
allows the derivation of the Newton's law of universal gravitation (\ref{eq10})
starting from first submicroscopic principles of the constitution of real
space \cite{20}. Submicroscopic mechanics further introduces
the correction (\ref{eq11}) to this law. This correction makes it possible to derive
exactly the same equations for the perihelion precession of Mercury, the
light deflection by Sun and the gravitational redshift of light
\cite{21}, which were derived by general relativity. Besides,
the submicroscopic concept uncovers inner reasons for the Shapiro time delay
(namely, the concept shows what exactly is hidden behind the fourth
component of the Schwarzschild metric) \cite{22}.

The submicroscopic concept introduces the tessel-lattice of mother-space as
a source and generator of matter and physics laws. The concept is fully
deterministic, removes an action at-a-distance and introduces a short-range
action, which is provided by photons in the area of electromagnetic
phenomena and inertons in the areas of quantum physics and gravitation. 
The concept complete rejects dark things from the space and inputs an additional elastic
interaction between gravitating objects caused by overlapping of object's
inerton waves (the notion of dark energy can be reduced to structural
peculiarities of the tessel-lattice at the universe scale, which requires a separate
consideration).

Although Zwitcky \cite{29} is treated as the ``father'' of dark matter concept,
the physical solution to this problem was demonstrated by Poincar\'{e}
\cite{30} even three decades before: describing the motion of an electron in
the ether Poincar\'{e} noted that the electron was surrounded by excitations
of the ether. Those excitations are interpreted as inertons of the
submicroscopic concept described in the present paper.

The submicroscopic concept allows us to launch a new project in astronomy,
namely, the Inerton Astronomy in the nearest future.

\end{document}